\documentclass[useAMS,usenatbib]{mn2e}
\usepackage{graphicx}
\usepackage{epstopdf}

\newcommand{\pasp} {Publications of the Astronomical Society of the Pacific}

\newcommand{\aj}{Astronomical Journal}
\newcommand{\apj}{Astrophysical Journal}
\newcommand{\apjl}{Astrophysical Journal, Letters}

\newcommand{\mnras}{Monthly Notices of the RAS}
\newcommand{\nat}{Nature}

\title[Spatial kinematics of BCGs]{Spatial kinematics of Brightest Cluster Galaxies and their close companions from Integral Field Unit spectroscopy\thanks{Based on VLT service
mode observations (Programme 381.B-0728) gathered at the European
Southern Observatory, Chile.}}
\author[S. Brough et al.]{S. Brough$^{1}$\thanks{E-mail: sb@aao.gov.au},
K.-V. Tran$^{2,3}$,
R. G. Sharp$^{1}$,
A. von der Linden$^{4}$,
Warrick J. Couch$^{5}$\\
$^{1}$Australian Astronomical Observatory, PO Box 296, Epping, NSW 1710, Australia\\
$^{2}$George P. and Cynthia W. Mitchell Institute for Fundamental Physics and Astronomy, Department of Physics and Astronomy,\\
Texas A\&M University, College Station, TX 77843\\
$^{3}$Institute for Theoretical Physics, University of  Z\"urich, CH-8057 Z\"urich, Switzerland\\
$^{4}$Kavli Institute of Particle Astrophysics and Cosmology (KIPAC), Stanford University, 452 Lomita Mall, Stanford, CA 94305\\
$^{5}$Centre for Astrophysics and Supercomputing, Swinburne University, PO Box 218, Hawthorn, VIC 3122, Australia\\
}
\begin{document}

\date{}

\pagerange{\pageref{firstpage}--\pageref{lastpage}} \pubyear{2011}

\maketitle

\label{firstpage}

\begin{abstract}
We present Integral Field Unit (IFU) spectroscopy of four brightest cluster galaxies (BCGs) at $z\sim0.1$.  Three of the BCGs have close companions within a projected radius of 20 kpc and one has no companion within that radius.  We calculate the dynamical masses of the BCGs and their companions to be $1.4\times10^{11}<M_{dyn} (M_{\odot})<1.5\times10^{12}$.  We estimate the probability that the companions of the BCGs are bound using the observed masses and velocity offsets. We show that the lowest mass companion (1:4) is not bound while the two nearly equal mass (1:1.45 and 1:1.25) companions are likely to merge with their host BCGs in 0.35 Gyr in major, dry mergers.  We conclude that some BCGs continue to grow from major merging even at $z\sim0$.  We analyse the stellar kinematics of these systems using the $\lambda_R$ parameter developed by the SAURON team.  This offers a new and unique means to measure the stellar angular momentum of BCGs and make a direct comparison to other early-type galaxies.  The BCGs and their companions have similar ellipticities to those of other early-type galaxies but are more massive.  We find that not all these massive galaxies have low $\lambda_{R_e}$ as one might expect.  One of the four BCGs and the two massive companions are found to be fast-rotating galaxies with high angular momentum, thereby providing a new test for models of galaxy evolution and the formation of Intra-Cluster Light.
\end{abstract}


\section{Introduction}

Brightest Cluster Galaxies (BCGs) include the most massive galaxies in the Universe. Models of hierarchical structure formation naturally feature the ongoing growth of the most massive galaxies by mergers \citep{peebles70}, and BCGs are predicted to have undergone more mergers than less massive galaxies (e.g. \citealt{delucia07}.  Thus
determining the merging history of BCGs is a particularly sensitive test of current formation models.  As the semi-analytic models of \cite{bower06,croton06} show, massive galaxies are over-produced in N-body dark matter cosmological simulations, and feedback mechanisms are required to bring the luminosity function into agreement with observations.  Tracing the recent assembly history of BCGs is vital to future development of galaxy formation models.


Observationally the evidence for BCG growth is contradictory: Studies of the luminosities and stellar masses of BCGs show little evolution in mass since $z\sim1-1.5$ (e.g. \citealt{brough02, collins09}) and their steep metallicity gradients are consistent with passive evolution since $z\sim2$ \citep{brough07}.  However, the large radii and low surface brightnesses 
of BCGs compared to normal elliptical galaxies are consistent with products of major, dissipationless mergers (e.g. \citealt{oegerle91,brough05, vDL07, lauer07}).  Their sizes and velocity dispersions may have also evolved faster than less-massive early-type galaxies since $z\sim0.3$ (\citealt{bernardi09} although, c.f. \citealt{stott11}).   
While BCGs are frequently observed to have multiple nuclei and close companions (e.g. \citealt{schneider83}), there are only a few examples where the companions have spectroscopically been confirmed to be bound (e.g. \citealt{tran08, rasmussen10}).  
BCG merger histories are also strongly linked to the formation of Intra-Cluster Light (ICL) as it is likely that a fraction of any galaxy merging with the BCG  ends up in the ICL (e.g. \citealt{conroy07,puchwein10}).

Integral Field Unit (IFU) spectroscopy has opened a new parameter space in which to analyse the stellar kinematics of early-type galaxies.  The SAURON team have developed a new parameter, $\lambda_R$, which utilises the increased spatial information from IFU spectroscopy to quantify the observed stellar angular momentum in galaxies \citep{emsellem07}: High-mass ($M_{\rm{dyn}}>2\times10^{11}~M_\odot$) early-type galaxies are dominated by slow rotators with low angular momentum, while lower mass early-type galaxies tend to be fast rotators with high angular momentum.  While this parameter provides a unique way to compare the stellar kinematics of BCGs to those of other early-type galaxies, the SAURON sample has only three galaxies with $M_{\rm{dyn}}>6\times10^{11}~M_\odot$ and only includes one BCG (M87).  Observations of BCG stellar kinematics to date have been limited to long-slit spectroscopy, (e.g. \citealt{kelson02}). The majority of BCGs do not show significant rotation in these observations, but a small number of massive BCGs are observed to undergo some rotation about their major-axes (e.g. \citealt{brough07, loubser08}) thus it remains unclear as to what angular momentum BCGs have.  

In this letter we present the first targetted observations of BCG stellar kinematics from IFU spectroscopy of three BCGs with close companions and one control BCG with no close companion.  We use the kinematic information to undertake a detailed analysis of whether the systems are likely to merge.  We then map the stellar kinematics of the BCGs and their companions and classify their angular momentum compared to the early-type galaxies of the SAURON sample.

Throughout this letter we assume a Hubble constant of $H_0=70$ km s$^{-1}$ Mpc$^{-1}$ and $\Omega_M=0.3$, $\Omega_\Lambda=0.7$.  

\section{Observations}

In \cite{vDL07}, a sample of 625 BCGs ($z\leq0.1$) were selected from the C4 cluster catalogue \citep{miller05} of the Third Data Release of the Sloan Digital Sky Survey (SDSS; \citealt{york00}). We have visually identified those BCGs with companions within $\sim10^{\prime\prime}$ (18 kpc at $z\sim0.1$).  Around 20 per cent of the sample have visually identified massive companions within $\sim10^{\prime\prime}$.  We observed three BCGs with companions, plus a fourth with no companion within the same radius as a `control' object.  The properties of the four host clusters are described in Table~\ref{tbl-1}.

\begin{table}
\begin{center}
\caption{Properties of clusters from von der Linden et al. (2007)}
\label{tbl-1}
\begin{tabular}{lccc c }
\hline
Cluster & RA &Dec & $\sigma$ & Log($M_{\rm{dyn}}$) \\
& && kms$^{-1}$ & $M_{\odot}$\\
\hline
1050 & 13:44:25.80 & +02:06:35.7 & 514 & 14.33 \\
1027 & 12:47:42.47 & -00:08:14.1& 1019 & 15.21\\ 
1066 & 13:31:10.83 & -01:43:48.9 & 814 & 14.92 \\
2086 & 23:22:56.37 & -10:02:44.1 & 599 & 14.52 \\
\hline
\end{tabular}
\end{center}

\end{table}

\begin{table*}
\begin{center}
\caption{Properties of BCGs from von der Linden et al. (2007), SDSS and IFU data. Log($M^{\star}_{~\rm{vdL}}$) is the stellar mass from von der Linden et al. (2007), measured from Petrosian magnitudes.  $z_{\rm{SDSS}}$, $R_{e,\rm{SDSS}}$ and $\epsilon_{\rm{SDSS}}$ are the redshift, effective radius (radius containing 50 per cent of the Petrosian flux) and ellipticity (measured from the flux-weighted second moments, Q and U) from the SDSS.  $z_{\rm{IFU}}$, $\lambda_{R_e, \rm{IFU}}$, $\sigma_{e, \rm{IFU}}$ and Log$(M_{\rm{dyn, IFU}}$) are the redshift, $\lambda$, velocity dispersion and dynamical mass measurements within $R_{e,\rm{SDSS}}$ from the IFU data. BCG 1027 has no redshift measured in SDSS.}
\label{tbl-2}
\begin{tabular}{lc c c c c c c c}
\hline
Galaxy & Log($M^{\star}_{~\rm{vdL}}$)& $z_{\rm{SDSS}}$ & $R_{e,\rm{SDSS}}$&$\epsilon_{\rm{SDSS}}$& $z_{\rm{IFU}}$ & $\lambda_{R_e, \rm{IFU}}$ & $\sigma_{e, \rm{IFU}}$ & Log$(M_{\rm{dyn, IFU}}$)\\
& $M_{\odot}$ & &$^{\prime\prime}$& $1-(b/a)$& && kms$^{-1}$&$M_\odot$\\
\hline
BCG 1050 &11.66 & 0.0721 & 6.29&  0.066&  0.0722  & 0.085$\pm0.005$&$399\pm2$  &  $12.20\pm0.55$\\
\hline
BCG 1027 & 11.34 & - & 4.64 & 0.093&  0.0894  & 0.104$\pm0.009$      &$263\pm2$   &  $11.80\pm0.29$\\
Comp 1027 &      -        & 0.0908   &  3.71        &0.035&  0.0909  &0.248$\pm0.004$&$ 224\pm4$ &  $11.57\pm0.20$\\
\hline
BCG 1066 & 11.36 & 0.0836 & 7.03 & 0.100&  0.0837  & 0.181$\pm0.006$ & $232 \pm3$ &  $11.84\pm0.26$\\
Comp 1066 &        -      & 0.0836   &  7.23        &0.312&  0.0836  &0.569$\pm0.006$  &$231\pm3$  &  $11.85\pm0.28$ \\
\hline
BCG 2086 & 11.25 & 0.0840 & 4.63 & 0.083& 0.0840  & 0.090$\pm0.006$ & $266 \pm8$ &  $11.78\pm0.46$\\
Comp 2086 &        -      & 0.0830   &  1.51        &0.024&  0.0819  &-  &$228\pm3$  &  $11.15\pm0.41$ \\

\hline
\end{tabular}
\end{center}

\end{table*}

The BCGs were observed in 2008 April and May with VIMOS.  
VIMOS was used in the IFU mode with the high-resolution, blue grism and a spatial sampling of
$0.67^{\prime\prime}$/pixel.  This gives a field-of-view of $27^{\prime\prime} \times 27^{\prime\prime}$ and a wavelength range of $4200-6200\rm{\AA}$, with spectral resolution of $2.1\rm{\AA}$ at $5100\rm{\AA}$ (corresponding to $\sigma_{inst}=52$ kms$^{-1}$).  Observations were made during dark time, with an average seeing of $0.7^{\prime\prime}$ (FWHM). Each BCG field was observed for $3\times1150$s.

Initial data reduction
was achieved using the \texttt{VIPGI} pipeline \citep{scodeggio05}.  The wavelength calibration is accurate to $0.09\rm{\AA}$ (from the $5577\rm{\AA}$ sky line).  Each observation consists of three dithered exposures split into four quadrants.  
For each quadrant, a sky background spectrum was calculated by taking the median over pixels without galaxy
light. 
The three exposures were combined with a $5\sigma$-clipped mean.  We used the 2-dimensional adaptive spatial binning code of \cite{cappellari03} to re-bin pixels to a minimum signal-to-noise ratio of 10 per pixel.

The stellar kinematics (velocity, V, and line-of-sight velocity dispersion, $\sigma$) of each galaxy were computed from the spectra of each bin, using a penalized pixel fitting scheme, (pPXF; \citealt{cappellari04}) 
and the MILES (Medium-resolution Isaac Newton Telescope Library of Empirical Spectra; \citealt{sanchezblazquez06}) evolutionary stellar population templates. These templates cover a similar wavelength range to VIMOS and have a similar spectral sampling (FWHM $= 1.8\rm{\AA}$).  We determined the optimal penalty value of 0.2  for pPXF by requiring the maximum bias in the GaussÐHermite parameters h3 and h4 to be equal to rms/3, where the rms is the scatter of the measurements obtained from Monte Carlo simulations with the adopted S/N and a well-resolved stellar dispersion $\sigma>180$ km s$^{?1}$.  The choice of penalty value does not significantly alter the result.  As an external check we compare the velocity measured in a central aperture of $3^{\prime\prime}$ for each galaxy with those from SDSS.  This gives a mean difference for the three BCGs and three companion galaxies with SDSS redshifts (Table~\ref{tbl-2}) of $cz_{IFU}-cz_{SDSS}=37 \pm 58$ kms$^{-1}$.


\begin{figure*}
\begin{center}	

\vspace{-1.cm}
\resizebox{45pc}{!}{
\hspace{-3.5cm}
\includegraphics{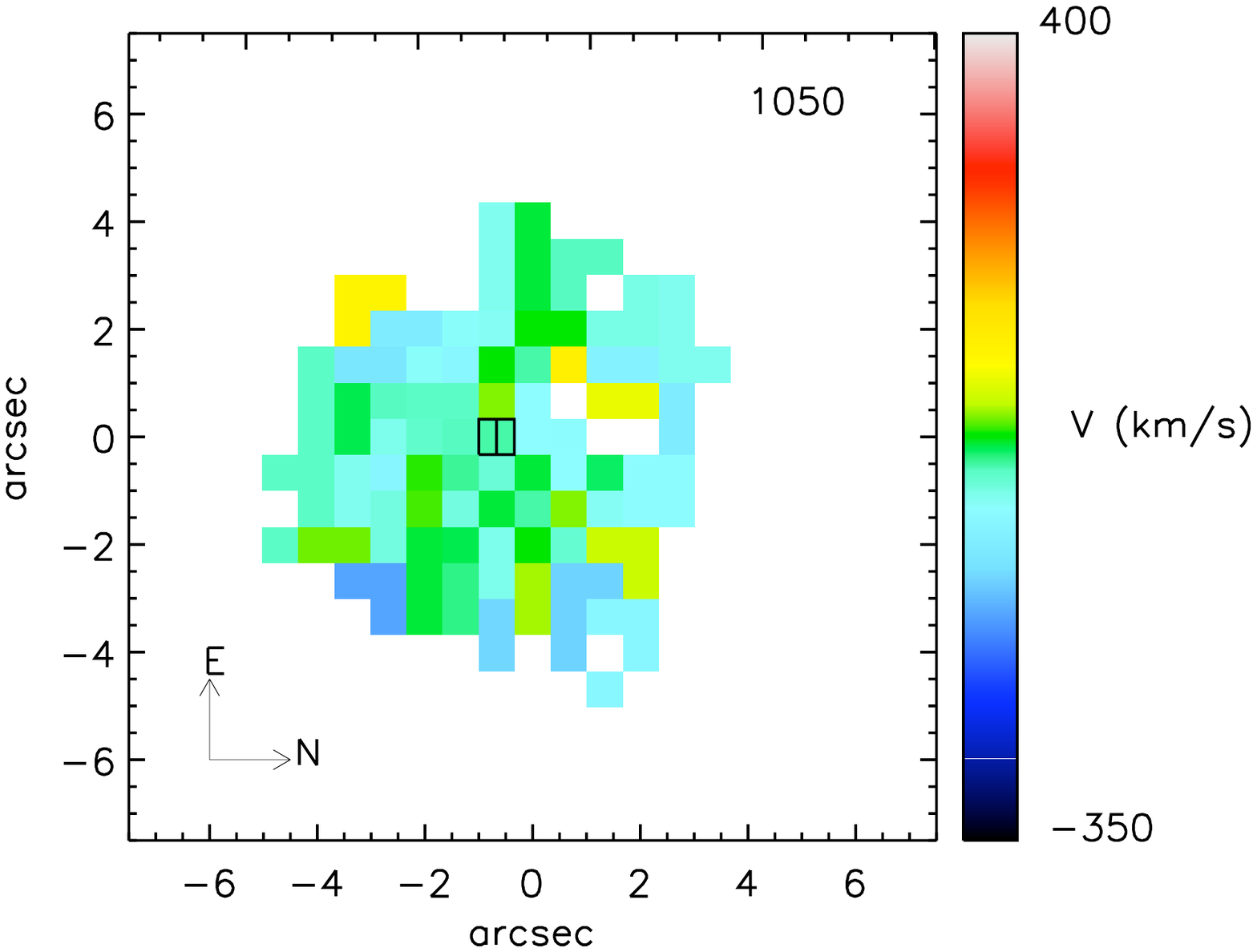}
\hspace{-3.5cm}
\includegraphics{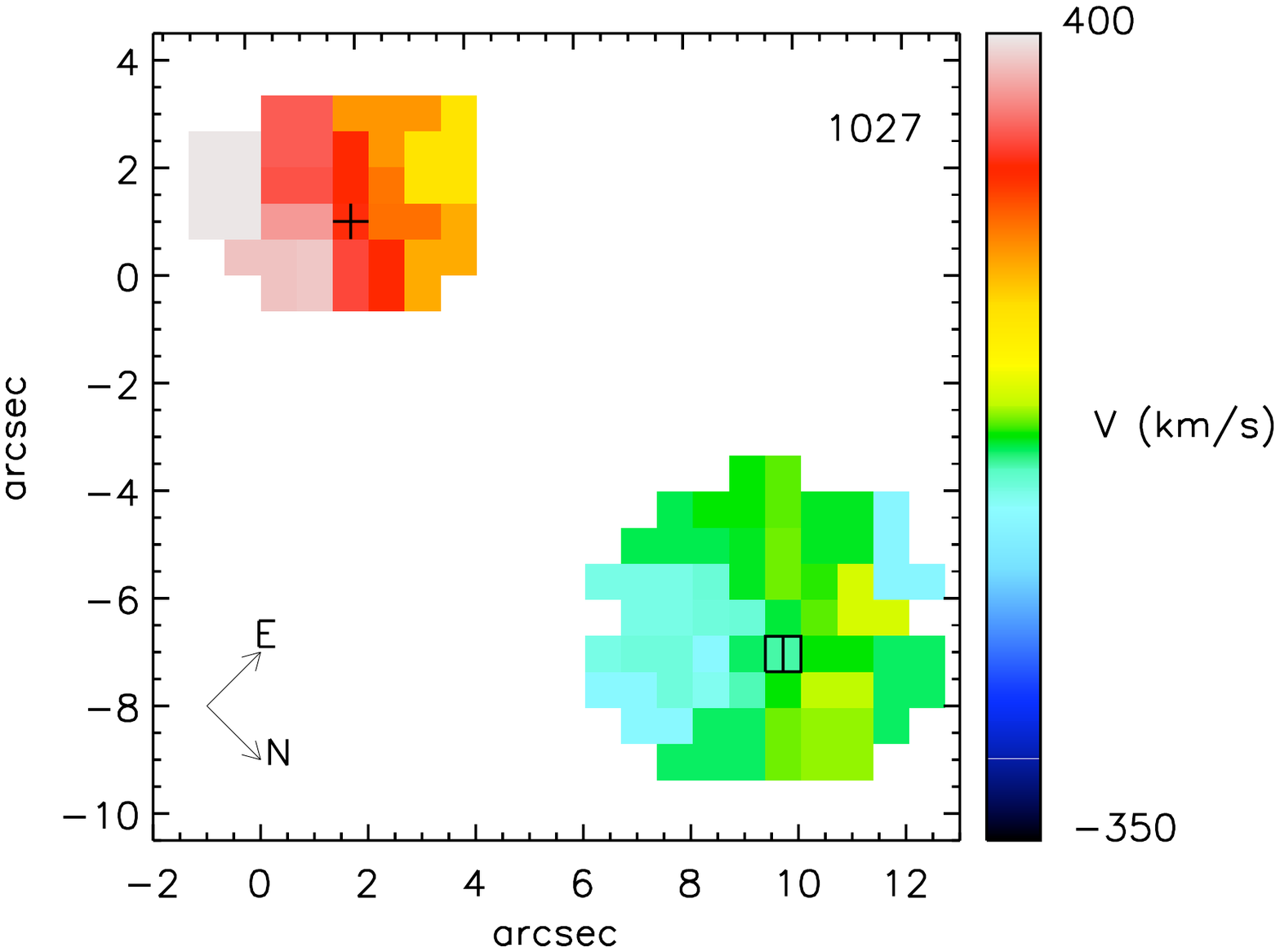}
\hspace{-3.5cm}
\includegraphics{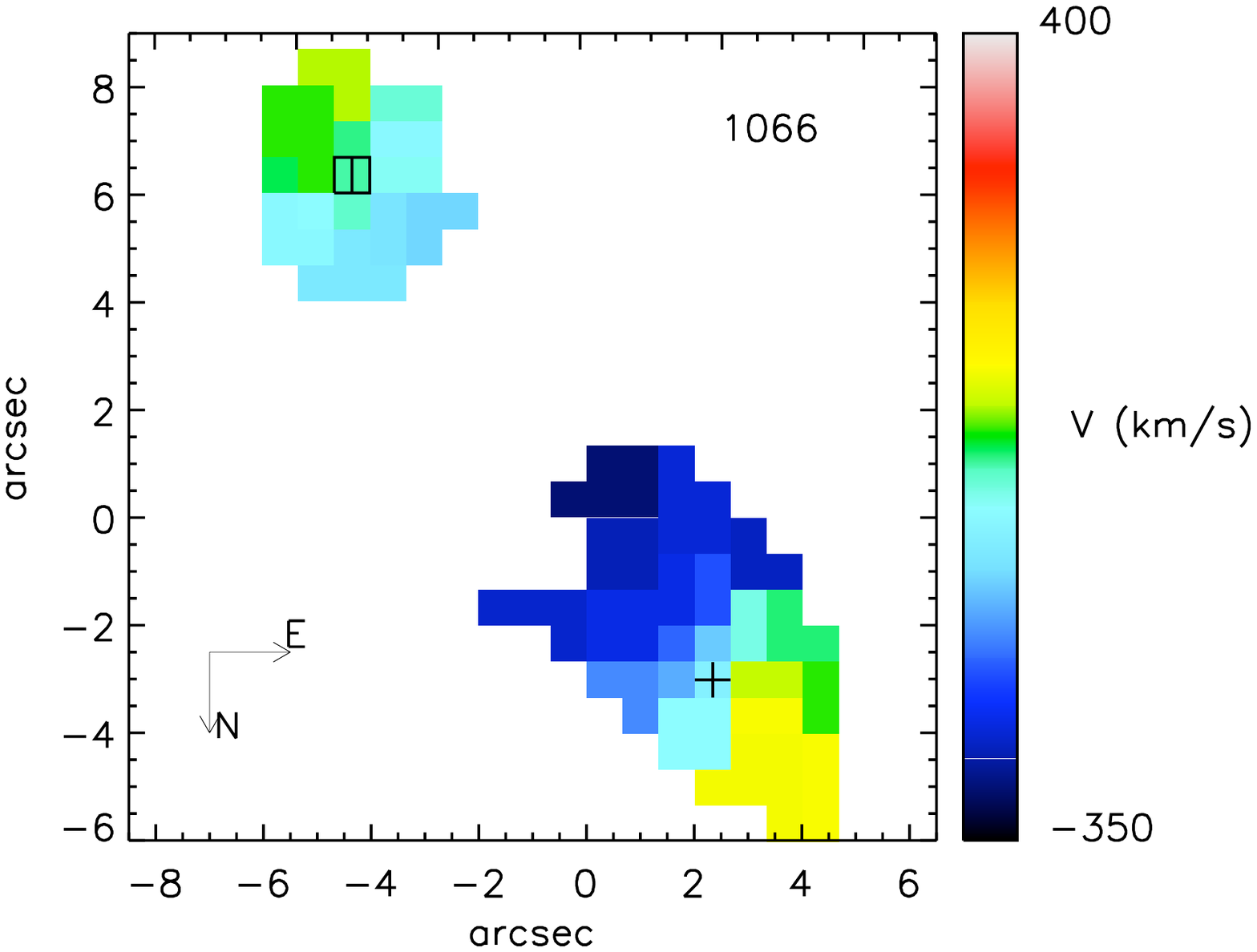}
\hspace{-3.5cm}
\includegraphics{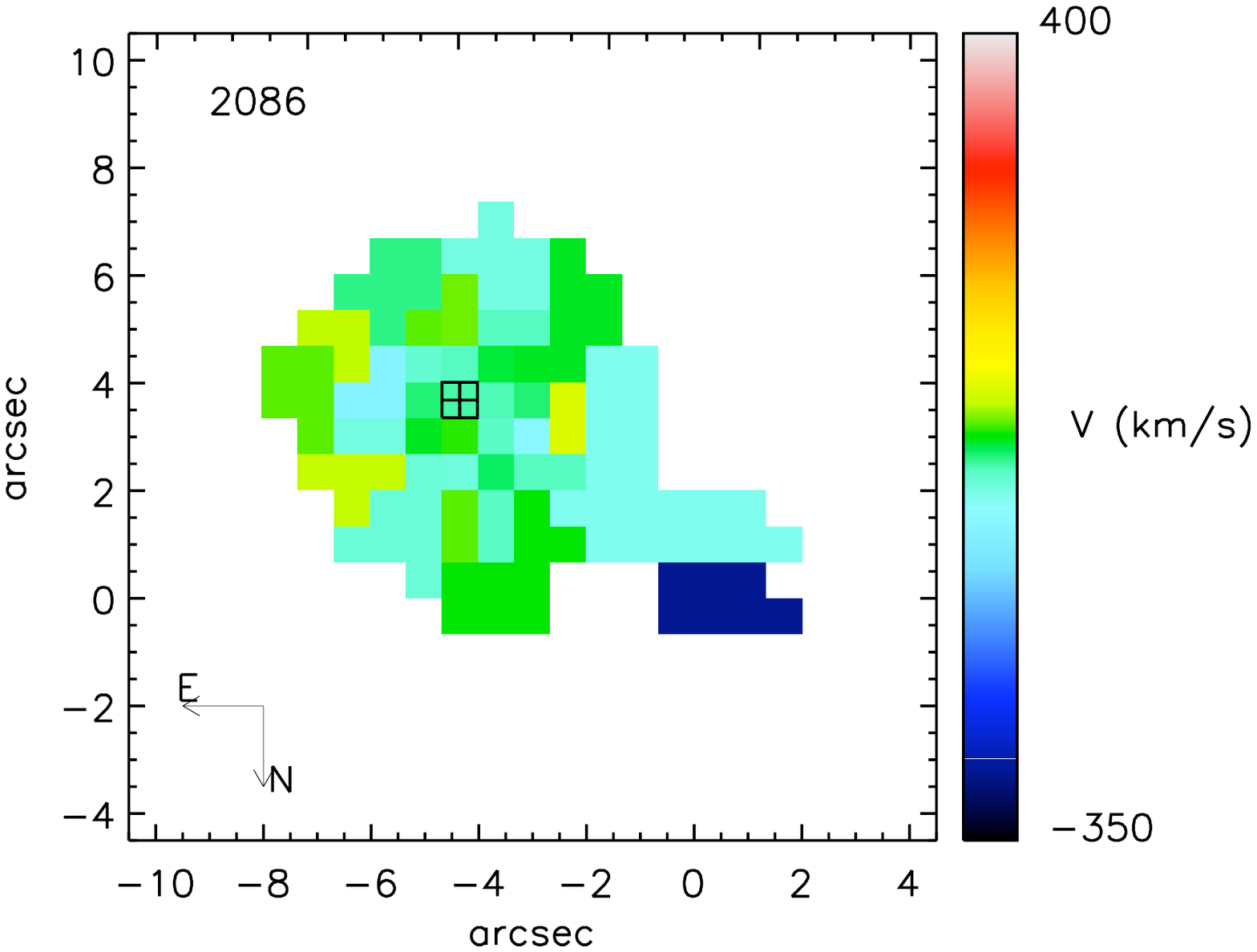}
}
\vspace{-3.5cm}
\end{center}
\caption{
Velocity maps scaled to $15^{\prime\prime} \times 15^{\prime\prime}$.  Left to right: Control BCG 1050, and BCGs 1027, 1066 and 2086.  The centres of the BCGs are indicated by squares and the companions by crosses. The orientation of the images is given by the arrows.  The BCGs in clusters 1027, 1050 and 2086 do not appear to be rotating, whilst the massive BCG companions and the BCG in cluster 1066 appear to rotate.}

\label{gals}
\end{figure*}

\section{Results}
\subsection{Are the companions bound?}

We first examine whether the companions are bound to the BCGs.  We estimate the probability of the systems being bound by considering the solid angles over which the systems would be bound, given the observed velocities and masses ($M_{\rm{dyn}}=5R_e\sigma_e^2/G$; \citealt{cappellari06}; Table~\ref{tbl-2}) and the Newtonian binding criterion that a two-body system is bound if the potential energy of the system is equal to or greater than the kinetic energy (e.g. \citealt{beers82,brough06}).

We first analyse the 2086 system which has a relatively low-mass companion.  The 2086 system is separated by a projected distance of 24 kpc ($2.2R_{e, \rm{BCG}}$), 565 kms$^{-1}$ and a mass ratio of 4:1.  This system is only likely to be bound at the 0.1 per cent level, suggesting that the low-mass companion is not bound to the BCG and is unlikely to merge.

Examining the systems with massive companions: The 1027 companion is at a projected distance of 18.1 kpc ($2.3R_{e, \rm{BCG}}$) and 261 kms$^{-1}$ from the BCG and has a mass ratio of 1.45:1.  This system is likely to be bound at the 61 per cent level, with a 99.993 per cent probability that the companion is on a bound inbound orbit.  The 1066 system is separated by a projected distance of 17.5 kpc ($1.6R_{e, \rm{BCG}}$) and 32 kms$^{-1}$ and a mass ratio of 1.25:1.  It is likely to be bound at the 98 per cent level with a similarly high 99.989 per cent probability that the companion is falling into the BCG.  These probabilities suggest that these systems are likely to merge.  
The 1027 and 1066 systems are observed to have luminosity residuals characteristic of interactions in the surface brightness analysis of C4 BCGs by Liu et al. (2009; using the data release 2 version of the C4 cluster catalogue these clusters are identified as 1026 and 1055).  In contrast, the 2086 system does not make their merger candidate sample.  Our detailed, kinematic analysis provides support for their photometric method of selecting merger candidates.  The dynamical friction formula of \cite{boylan-kolchin08} suggests that, depending on the eccentricity of the orbit, these systems would take between 0.2 and 0.35 Gyr to merge, consistent with estimates made using a variety of methods by \cite{liu09}.  There are no emission lines in our spectra so 
these are likely to be dry mergers.  The companion galaxies also have very high dynamical masses, meaning that if they merge they will undergo nearly equal-mass major mergers. 

This is concrete evidence that some BCGs continue to grow by major, dry, mergers, even at $z\sim0$.
 
 \subsection{Stellar Kinematics}
We present two-dimensional velocity maps in Figure~\ref{gals}. The lack of a well-ordered velocity gradient across the face of BCGs 1027, 1050 and 2086 suggest that they do not rotate, as would be expected for massive early-type galaxies.  In contrast, BCG 1066 and the two massive companion galaxies appear to be rotating. The low signal-to-noise of the BCG 2086 companion precludes detailed kinematical analysis.

We quantify the angular momentum of these systems and how that relates to other early-type galaxies using the SAURON $\lambda_R$ parameter \citep{emsellem07}. $\lambda_R$ was developed as a proxy for the observed projected stellar angular momentum per unit mass:
$\lambda_R = \langle R|V| \rangle / \langle R \sqrt(V^2 + \sigma^2) \rangle$.
$R$ is the observed distance to the centre of the galaxy, $V$ the velocity, and $\sigma$ the line-of-sight
velocity dispersion.  The brackets correspond to a flux-weighted sky average.  $\lambda_R$ tends to unity when the mean stellar rotation dominates. We estimate the error in $\lambda_R$ through a 10,000 iteration Monte Carlo simulation of the measured errors in $V$ and $\sigma$ on $\lambda_R$ and find the error to be of the order of a few per cent.

\begin{figure}
\begin{center}	
\resizebox{15pc}{!}{
\includegraphics{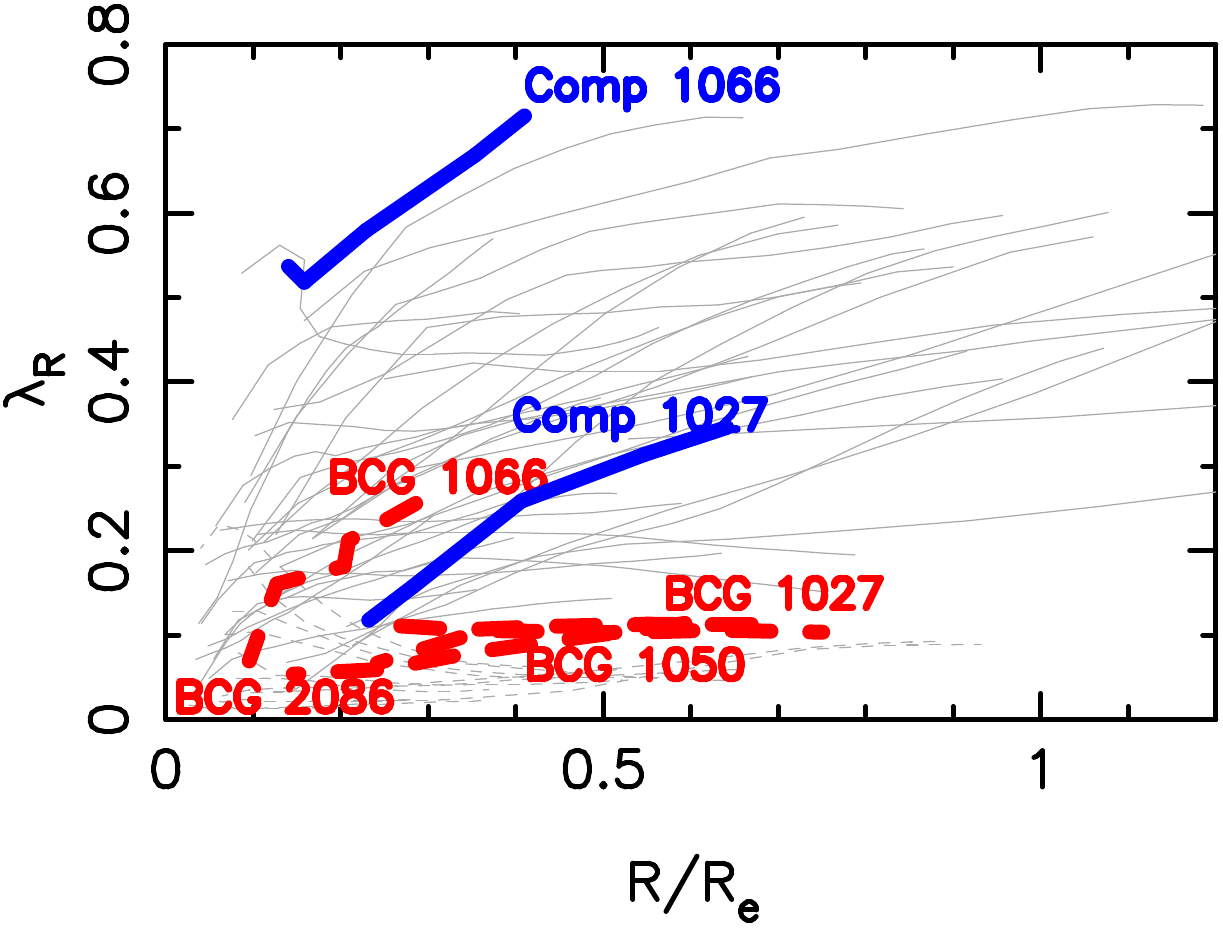}
}
\end{center}
\caption{Radial $\lambda_R$ profile for the 4 BCGs (thick red dashed lines) and 2 massive companion galaxies (thick blue lines).  The 48 early-type galaxies of the SAURON sample are shown in grey thin lines with fast (slow) rotators shown by the solid (dashed) lines.  BCGs 1027, 1050 and 2086 are consistent with the slow rotators of the SAURON sample whilst BCG 1066 and the companion galaxies are consistent with the fast rotators.}
\label{Sfig4}
\end{figure}

We show $\lambda_R$ profiles for our galaxies compared to the 48 SAURON early-type galaxies in Figure~\ref{Sfig4}.  The profiles of BCGs 1027, 1050 and 2086 are consistent with the slow rotators of the SAURON sample while BCG 1066 and the massive companion galaxies are consistent with the fast rotators.  We note that the data for BCG 1066 only reach a small fraction of its effective radius and this profile could turn over at larger radii.  However, the steep profile is unlikely to be due to a kinematically distinct core (KDC) as these are generally $\sim$kpc sized \citep{emsellem07} and would only be visible within $0.1~R_e$ for BCG 1066.

\begin{figure*}
\begin{center}	
\resizebox{33pc}{!}{

\includegraphics{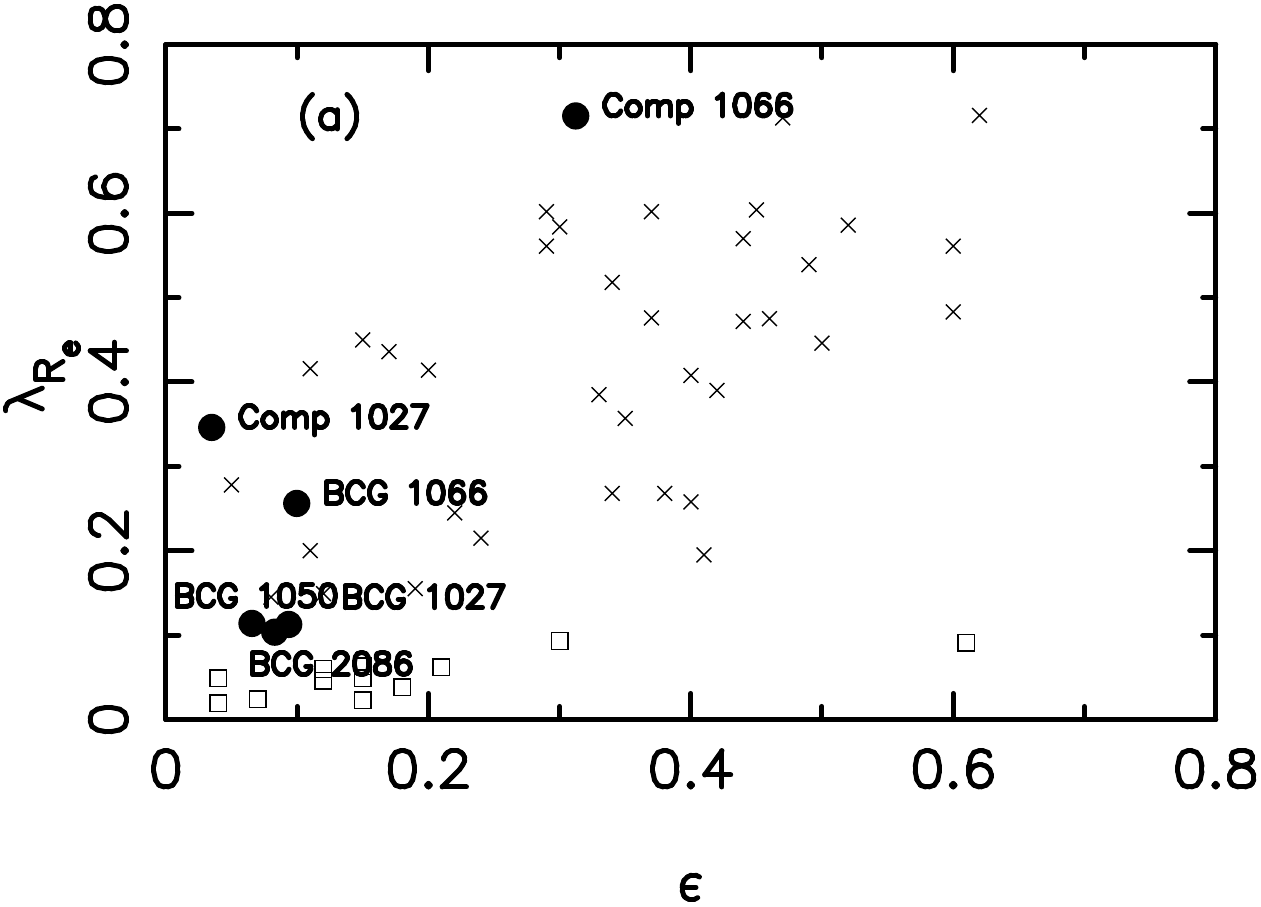}
\includegraphics{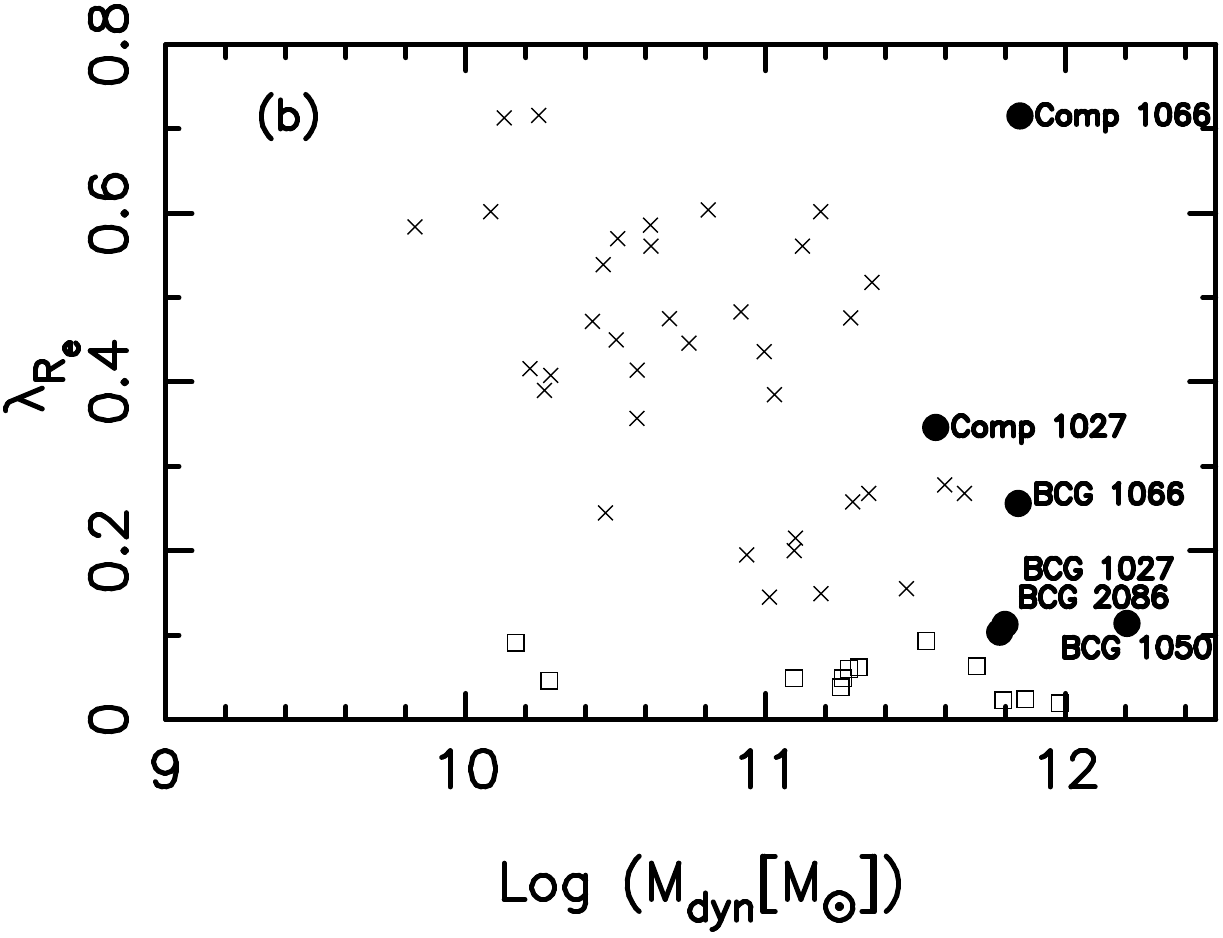}

}
\end{center}
\caption{$\lambda_{R_e}$ versus (a) ellipticity, $\epsilon$, and (b) dynamical mass, $M_{\rm{dyn}}$, for the 4 BCGs and 2 massive companion galaxies (solid points) with the typical uncertainty given by the error bar. The 48 early-type galaxies of the SAURON sample are divided into fast rotators (crosses) and slow rotators (squares).  BCGs 1027, 1050 and 2086 are consistent with the SAURON slow rotators whilst BCG 1066 and the companion galaxies are consistent with the fast rotators.  The BCGs and their companions are rounder and more massive than the majority of the SAURON sample.}
\label{Sfig5}
\end{figure*}

We take a measure of $\lambda_R$ at the effective radius, $\lambda_{R_e}$, or the largest radius in our data (c.f. \citealt{emsellem07}; Table~\ref{tbl-2}).   BCGs 1027, 1050 and 2086 have similar $\lambda_{R_e}$ values. In contrast, BCG 1066 and the massive companions have significantly higher values. Figure~\ref{Sfig5} illustrates where our systems lie with respect to the SAURON sample in terms of their ellipticity (from SDSS) and dynamical mass.  The BCGs have low ellipticities compared to the SAURON galaxies.  We find them to have high dynamical masses showing that we are probing the massive end of the galaxy distribution. The low $\lambda_{R_e}$ of control BCG 1050 and BCGs 1027 and 2086 is consistent with the slow rotating SAURON galaxies while the high $\lambda_{R_e}$ of BCG 1066 and the companion galaxies is consistent with the fast rotating SAURON sample.

\section{Discussion}
\label{discuss}

We observe here two BCGs at $z\sim0.1$ that are undergoing major mergers.  This is evidence that some BCGs grow via major dissipationless merging even at $z\sim0$.  We also show that one BCG with a close companion in projection is not bound, demonstrating the need for kinematic analysis.

These observations are in contrast to studies that claim BCGs have not increased in mass since $z\sim1$ (e.g. \citealt{whiley08,collins09, stott10}).   
However, high-resolution hydrodynamical simulations predict that in a major merger with a BCG, not all the companion galaxy merges with the BCG: 50-80 per cent of its mass will join the ICL \citep{conroy07,puchwein10}.  We therefore infer that most of the stellar mass from late major merging goes into the ICL.  This is consistent with observational measurements of the ICL by \cite{gonzalez05}.

With IFU observations we open a new parameter space with which to test BCG formation models.  Our very massive BCGs complement the SAURON results (in which there are only three massive galaxies $M_{\rm{dyn}}>6\times10^{11}~M_\odot$; \citealt{emsellem07}; Figure~\ref{Sfig5}) and the upcoming ATLAS3D results which, being a volume-limited sample, has limited numbers of high-mass galaxies \citep{cappellari10}. We show that not all massive BCGs have low angular momentum as one might expect.  In fact, one out of our four BCGs is a fast rotator.  We speculate here that the angular momentum of the BCG with the highest $\lambda_{R_e}$ is related to its interaction with a very close, massive merging companion.  However, we do not observe a similar increase in the 1027 system which also has a close, massive companion so the rotation of BCG 1066 could be intrinsic.  The two massive companion galaxies also show evidence for rotation. 



This investigation serves as a pilot study. In the future we intend to investigate a larger sample of BCGs with close companions to further examine whether they are bound, and to test the hypothesis that higher $\lambda_{R_e}$ in massive galaxies could be an indication of ongoing/recent merging.  We also intend to compare directly to the high resolution ($<1$ kpc) hydrodynamical cosmological simulations of a Virgo-like galaxy cluster of \cite{teyssier10} to further understand these results from a theoretical perspective.  



\section{Conclusions}

We have analysed the first observations of stellar kinematics from Integral Field Unit (IFU) spectroscopy of four brightest cluster galaxies (BCGs) at $z\sim0.1$.  Three of the BCGs have companions within a projected radius of 20 kpc and one has no companion within that radius.  

Undertaking a detailed analysis of whether the companions are bound to the BCGs we find new evidence that some BCGs grow via major dissipationless merging at $z\sim0$, with two of the three multiple systems in our sample having a high probability that they are dynamically bound and likely to merge within $\sim0.3$ Gyr.  For BCGs to undergo mergers and also be consistent with studies that claim no evolution in the stellar mass of BCGs since $z\sim1$ (e.g. \citealt{whiley08, stott10}) suggests that most of the stellar mass from the late merging goes into the Intra-Cluster Light.

The $\lambda_{R_e}$ parameter developed by the SAURON team offers a new and unique means to measure the projected stellar angular momentum per unit mass of galaxies.  It enables a direct comparison of the stellar kinematics of BCGs to those of other early-type galaxies.  The BCGs and their companions have similar ellipticities to those of other early-type galaxies in the SAURON sample but they are more massive.  We find that not all BCGs have low angular momentum as might be expected from an extrapolation of the SAURON sample. One of our three multiple-nuclei BCGs is a fast rotator, as are two of their similarly massive companions.  These observations provide a new test for modeling how the most massive galaxies in the universe form.

\section*{Acknowledgments}

We thank the anonymous referee for helpful comments.  We thank Eric Emsellem for providing guidance and the SAURON data.  The data published in this paper have been reduced using VIPGI, designed by the VIMOS Consortium and developed by INAF Milano.


\bsp

\label{lastpage}

\end{document}